\documentclass[12pt]{article}
\usepackage{graphicx}
\begin{document}
\title{Structure and light emission of a Str\"omgren system.}

\author{Jacques Moret-Bailly.}

\maketitle
mail: jmo@laposte.net\begin{abstract}
Stroemgren defined a model made up of an extremely hot source plunged in a constant density, huge, static cloud of low pressure hydrogen. The present studies of this model apply qualitatively with relatively small and cold sources, but without these assumptions, we must take into account that: - the source has, at all frequencies, the spectral radiance of a laser at a single frequency, so that the multi-photon absorption of few lines involves the whole continuous spectrum; - a spherical shell containing large column densities of excited atoms emits strongly super-radiant beams selected by competition of modes; - the long paths in excited atoms allow observations which require in the labs the use of ultrashort laser pulses.  A necessary, qualitative update watch that: - the whole continuous spectrum of the source pumps the atoms to excited states, so that almost all energy emitted by the source is transferred to a line spectrum; - the main fraction of this line spectrum is emitted by a strong super-radiance in a spherical shell where the temperature becomes cold enough for a notable de-ionization of the atoms; by competition of modes, a dotted ring appears; - inside the ring, a less bright region emits spontaneously lines made extremely broad by parametric transfers of energy between the light beams and thermal radiation; the intensities of these lines decrease down to zero, from increased laboratory frequencies to decreased frequencies. Supernova remnant 1987A shows these properties.
\end{abstract}

%{270.6630, 270.4180, 270.1670.} 
super-radiance, multi-photon processes, coherent optical effects.

\section{Introduction}\label{intro}
Str\"omgren \cite{Stromgren} studied, in a stationary state, the structure and the spectroscopy of an isotropic model made up of a source  hot enough to emit much vacuum UV, surrounded by a large, constant density, static cloud  of low pressure hydrogen, cold at a long distance. Later work on this model, founded on the assumption of a local re-absorption of the lines of neutral hydrogen atoms (assumption \textquotedblleft on the spot\textquotedblright), or criticizing this assumption \cite{Ritzerveld}, did not qualitatively modify the properties of the model. 

\medskip
A Str\"omgren system depends on several parameters: size and temperature of the source, density of hydrogen; the aim of the present improvement of the model is the introduction of {\it new qualitative properties}, opening the way to particular, numerical studies.

\medskip
Up to now, the model was studied with the implicit hypothesis of linear, incoherent interactions of light with matter. Here, supposing that the temperatures of beams\footnote{To apply thermodynamics more easily, the use of temperatures of beams, deduced from their energetic, spectral radiances and their frequencies by Planck's law, is preferred to the use of spectral radiances.} radiated by the source are larger than $10^6$\ K, and that the excited medium is large, three types of interactions requiring methods of laser spectroscopy must be used:

i) Super-radiance was observed initially by the sharpening of spectral lines emitted along the axis of discharge tubes, and by the appearance of a Dicke spike. In this super-radiance, qualified \textquotedblleft weak\textquotedblright, the induced emission remains lower than spontaneous emission. In \textquotedblleft strong super-radiances\textquotedblright{} spontaneous emissions are lower than induced emissions. The emission induced by, and amplifying a super-radiant beam strongly extracts energy from the excited atoms, up to an equilibrium of the temperature of the beam with the temperature of the transition (deduced from Boltzman's law), unless the populations of the involved levels are inverted. By the decrease of the population of the higher level, the spontaneous emission and the amplification of weaker beams become low, so that this \textquotedblleft competition of modes\textquotedblright{} limits the number of super-radiant beams. The remaining beams correspond to paths for which the initial amplification is maximal.
A high super-radiance works in the lasers, the optical paths in the amplifying medium being virtually increased by a cavity for its resonance wavelengths. The cavity memorizes a phase, so that a laser emission is more time-coherent than a simple super-radiant emission.

ii)  Suppose that the state of an atom is changed several times by a set of successive interactions. The interval of time between the centers of two successive interactions is short if the radiances of the exciting light beams are large. This interval of time may be shorter than the duration of an interaction, so that the interactions are mixed, bound into a multi-photon interaction. During the set of interactions, the atom is \textquotedblleft dressed\textquotedblright{} by the electromagnetic fields in a non-stationary state, a time-dependant mixture of stationary states, in which the initial then final states are preponderant. Some intermediate states may be virtual. In the labs, the multi-photon interactions require the use of lasers, but an extremely hot source has a similar radiance in its whole spectrum.

iii) A parametric effect results from space-coherent interactions of several light beams within a medium, so that the medium, dressed by the fields during the interaction, returns to its initial state after the interaction: The medium plays the role of a catalyst. Preserving the space-coherence requires an identical relation of the phases of the beams at each atom; this may be obtained using several indexes of refraction of a crystal, so that the wavelength may be equal at different frequencies; this method is used to multiply, combine,... frequencies of laser beams. G. L. Lamb Jr. \cite{Lamb} shows that the space-coherence may be preserved by a decrease of the time-coherence through the use of \textquotedblleft ultra-short\textquotedblright{} light pulses. Ultra-short is not a very convenient word, because the property \textquotedblleft ultra-short\textquotedblright{} implies the medium, an ultra-short pulse being defined by Lamb as \textquotedblleft shorter than all relevant time constants\textquotedblright{}. An example is the decrease of the frequencies of the light pulses sent into optical fibers to transmit information: This decrease is proportional to the length of the fibers. This disturbance results from a coherent transfer of energy to the thermal background.

Section \ref{str} shortly presents the original Str\"omgren system, in which a relatively thin shell of a plasma mixes neutral atomic hydrogen H$_I$ and ionized hydrogen H$_{II}$. In this \textquotedblleft Str\"omgren shell\textquotedblright, the radiation of the highly excited atoms is large. This shell surrounds a \textquotedblleft Str\"omgren sphere\textquotedblright{} mainly made of free protons and electrons.

Section \ref{sura} introduces the super-radiant emission of this spherical shell of plasma.

Section \ref{insca} shows how nearly all light emitted by the source amplifies the super-radiant beams.

Section \ref{shift} and the appendix show how to replace, in the theory of refraction, coherent Rayleigh scattering by coherent Raman scattering, to obtain frequency shifts.

Section \ref{spec} studies the spectrum of light spontaneously emitted inside the Str\"omgren sphere.

Figure 1 shows  the evolution, versus the distance $r$ to the source, of the local parameters: state of hydrogen, temperature, nature and properties of light. The progressive building of this figure is described in the text which may then use the built part of the of figure for its demonstrations.

Figure 2 shows the shape of the lines emitted spontaneously in the sphere.

\medskip
{\it Sections  \ref{sura} and \ref{insca} improve Str\"omgren's results in the hypothesis that the column densities of excited atomic hydrogen in the  \textquotedblleft Str\"omgren shell\textquotedblright{} are large enough to start a strong super-radiance; the frequency shifts described in sections  \ref{shift} and \ref{spec} work in any system containing excited atomic hydrogen.}

\section{The Str\"omgren model.}\label{str} % 2
Close to the source $O$, hydrogen is fully ionized into protons and electrons which, at a low pressure and in a supposed negligible magnetic field, are nearly free, therefore almost without interaction with light. Consequently, the transfers of energy result mainly from the diffusion of the atoms, so that the temperature is a decreasing funct\medskip
Study the total amplification coefficients $k(\rho)$ of parallel rays lying in a plane $\Pi$ containing $O$, according to their distances $\rho$ to $O$. To study this amplification, split the regions containing H$_I$*, that is the Str\"omgren shell and the external region of the sphere, into infinitesimal shells $S$ centred at $O$. If $\rho$ is small, the ray crosses all shells $S$ twice, almost perpendicularly. Increasing slightly $\rho$, the ray crosses the $S$ under a smaller angle, that is along a larger distance. Thus, the ray is more amplified, so that $k(\rho)$ is an increasing function of $\rho$. If $\rho$ is large, so that the ray does not cross the shell, $k(\rho)=0$. Thus, $k(\rho)$ has a maximum $k(R')$. 
ion of the distance from the source $r$.

However, after a decrease of temperature around 200\ 000\ K, colliding electrons and protons may be considered as atoms in a collisional state, able to radiate some energy, so that the temperature decreases faster and faster versus $r$.
Assuming the spherical symmetry, a  \textquotedblleft Str\"omgren sphere\textquotedblright{} of radius $R$ is made of almost fully ionized hydrogen (graph 1:b). Around $10^5$\ K some excited atoms appear in the outer region of the sphere. They cool by emission of light, becoming able to catalyze the formation of new atoms by their reactive collisions with protons and electrons. Thus, the formation of atoms self-accelerates, nearly exponentially, while the emission of spectral lines of hydrogen increases. These evolutions are represented on the left regions ($r<R$) of graphs 1:a and 1:c. 

The sphere is surrounded, for $r$ increasing from $R$ to $R1$, by a plasma of atoms, protons and electrons, the \textquotedblleft Str\"omgren shell\textquotedblright{} which radiates a lot, so that it cools fast and is relatively thin. Out of this shell, hydrogen is atomic, in its ground state, then molecular. In the shell, the temperature decreases, roughly, from 30\ 000\ K, to a temperature at which the neutral atoms are stable. At the limit $R1$, the ionization and excitation disappear because the far UV emitted by $O$ was fully absorbed by the shell.

\begin{figure}
\centerline{\includegraphics[height=12 cm]{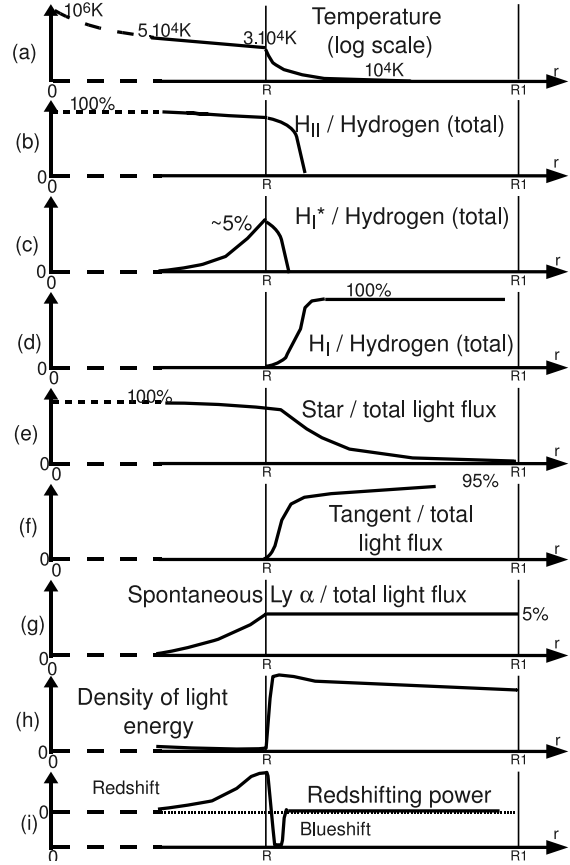}}
\caption{Orders of magnitude of the variations along a radius of: (a) the temperature of hydrogen; (b): the proportion of ionized hydrogen H$_II$; (c): the proportion of excited atoms; (d): the proportion of unexcited atoms; e): the proportion of energy of light directly emitted by the source; (f): the proportion of energy transmitted to the super-radiant tangential rays; (g): the proportion of the energy transmitted to the spontaneous emission; (h) the density of luminous energy which is strongly increased where the direction of propagation is turned of $\pi/2$; (i) the red-shifting power for low energy spontaneous emission; for $r>0$, its modulus is nearly proportional to the density of H$_I$* (c) and it depends on the other beams. See text. }
\label{rad1}
\end{figure}

\medskip
Str\"omgren and following authors implicitly assumed that, in the shell, all emissions of light by the excited atoms (represented by H$_I$*) are incoherent, neglecting induced emissions. They suppose that the emissions are followed by ``on the spot'' re-absorptions takink no account of the phases. Using such computations, many physicists said: \textquotedblleft Townes maser will not work\textquotedblright. Emission and absorption along a light beam must be studied simultaneously using Einstein coefficients A and B or only B if the zero point field is included in the radiance. 

 We will show, in the next section, that, for $r>R$ the cooling becomes faster than found by Str\"omgren and set a more precise value of $R$.

\section{Super-radiant emission.}\label{sura} 
The Str\"omgren shell $(R<r<R1)$ contains a plasma in which atoms are excited.  This medium amplifies strongly the lines of hydrogen (it is a gas laser plasma). The start of the qualitative difference between our model and Str\"omgren's is: We {\it assume} that the power of the source generates in a large enough cloud of hydrogen a volume of plasma in which the column density of excited atoms is large enough to start a strong super-radiant emission. We define $R$ as the distance at which this emission starts. With the assumption of an extremely powerful source and a huge cloud of hydrogen, there is necessarily a region where the integral of the density of excited atoms along some paths (\textquotedblleft column density\textquotedblright{}) is as large as the product of the column density by the factor of quality of the cavity in a laser. On graph 1:c, to give an order of magnitude, we write that, as at the start of many gas lasers, the proportion of excited atoms is of the order of $5\%$.

A super-radiant ray induces emissions which amplify it, depopulating the excited levels of the gas.  This depopulation is strong where the radiance of the beams (graph 1:f) is large. As the production of H$_I$* from of H$_{II}$ is limited by the long time between collisions at low pressure, the density of excited atoms becomes negligible (graph 1:c) although it remains some H$_{II}$ (graph.  1:b).

 The super-radiances reduce the lifetime of all excited states, including the ionization states, that is the collisional states of a proton and an electron; thus, elastic collisions may become inelastic and produce directly atoms in their ground states. In this process, the transitions between high energy states have low energies, they produce much thermal radiation. As in many gas lasers, the amplification of light is not limited by the speed of de-excitation of the atoms but by the speed of generation of high states, here collisional states of protons and electrons.

\medskip
Study the total amplification coefficients $k(\rho)$ of parallel rays lying in a plane $\Pi$ containing $O$, according to their distances $\rho$ to $O$. To study this amplification, split the regions containing H$_I$*, that is the Str\"omgren shell and the external region of the sphere, into infinitesimal shells $S$ centred at $O$. If $\rho$ is small, the ray crosses all shells $S$ twice, almost perpendicularly. Increasing slightly $\rho$, the ray crosses the $S$ under a smaller angle, that is along a larger distance. Thus, the ray is more amplified, so that $k(\rho)$ is an increasing function of $\rho$. If $\rho$ is large, so that the ray does not cross the shell, $k(\rho)=0$. Thus, $k(\rho)$ has a maximum $k(R')$. 

The spherical symmetry is broken by the competition of the modes, so that $k(R')$ may have many values. The largest values $k(R)$ of the $k(R')$ correspond to the superradiant rays, defining $R$.

The break of the spherical symmetry may be initiated by stable fluctuations of the system, so that the system of modes becomes stable.

\medskip
Straight lines tangent to a sphere and parallel to a given direction generate a cylinder. Thus the super-radiant rays observed from far are on a cylinder of radius $R$ tangent to the sphere on a circle. These rays, enlarged by diffraction, are axis of modes. Their competition selects some of them. Thus the circle is seen as a regularly dotted ring similar to the dotted ring showed by transverse modes TEM($l,m$) of a laser, for which a selection of modes fixes the radial parameter\ $l$.

The super-radiance cools the gas more than the spontaneous emission assumed by Str\"omgren, obtaining a negligible ionization at a lower distance R1. We now suppose that figure 1 is adjusted for this smaller value of $R1$.

\medskip
Set p, q, u, v, labels of increasing energy levels.  If the transition from u to q is super-radiant, the increase of population of q favors emissions from q to p, the decrease of population of u favors emissions from v to u. Thus, there is a large probability to obtain the emission of the whole cascade from v to p, for H$_I$ of the set of alpha emissions. As the super-radiance is strong, the life times of all excited states are short, so that the de-excitations tend to be simultaneous, in a multi-photon process. The correlation of emissions of a cascade works along the paths of the super-radiant rays which become polychromatic. But, as the diffractions differ and the amplification coefficients decrease as a function of $r$, the distances of the axis of the modes of higher frequencies are closer to\ $R$. Thus, if the frequencies are not selected in the observation, the dots appear brighter at their inner rim.

\section{Induced scattering.}\label{insca}
Abundant hydrogen atoms  H$_I$  appear in the ground state as a consequence of super-radiant, induced tangential emission (graph 1:d) which de-excite the atoms  H$_I$*, and, as fast as protons and electrons collide, de-excite H$_{II}$ (graph 1:b).

An atom may be excited by several laser beams whose combination of frequencies corresponds to a transition; it is a multi-photon excitation. The atoms of hydrogen may be excited by multi-photon excitations resulting from a simultaneous interaction with several frequencies of the continuous spectrum of the source (which have the same spectral radiance than a laser). Thus, with various combinations of frequencies, the whole continuous spectrum of the source may pump the hydrogen atoms to high states.

\medskip
Strong emissions from high states may be induced by the super-radiant beams.

A strong induced emission combines with the multi-photonic absorption of the radial beams into a \textquotedblleft multi-photon induced scattering\textquotedblright{} which transfers energy from the radial continuous spectrum to the line spectrum of the atoms; on graph 1:e, we represent that the absorption of the flux of the star increases much where the number of atoms increases. In few words, the {\it extreme} temperature of the radial rays at all frequencies, and the {\it relatively high} temperature of the tangential rays at the resonance frequencies, rub out obstacles to a thermodynamic quasi-equilibrium of these rays at their interacting frequencies.

A large fraction of the energy radiated by the source is transferred to the super-radiant beams. The spectral radiance of the superradiant rays tends to the spectral radiance of the source. Although the total radiance of the source remains larger than the total radiance of the super-radiant beams, if the source is observed through a solid angle much smaller than the solid angle of observation of the ring, the remaining flux of the source is radically lowered, the source is invisible.

In the cold gas which surrounds the shell, a cylindrical, mono mode, amplified super-radiant beam may excite a cylinder of impurities, cold atoms or molecules, making a long cylinder of excited particles able to emit super-radiant, very sharp lines into the same direction.

\section{Propagation of light in low pressure excited hydrogen.}\label{shift}
While the Rayleigh coherent scattering of light, named refraction, is a fundamental interaction of light with matter, observing other coherent effects is difficult because they must be parametric to avoid an excitation of the medium, and because preserving the coherence requires {\it a priori} the same wavelength at various frequencies: However G. L. Lamb describes the use of \textquotedblleft ultra-short light pulses" \cite{Lamb}: the light pulses must be \textquotedblleft shorter than all relevant time constants". With a convenient matter, ordinary, incoherent light, made of nanosecond pulses, 10$^5$ times longer than the usual femtosecond pulses, appears made of ultra-short pulses.

To verify Lamb's condition using ordinary incoherent light, the collisional time and a quadrupolar resonance period must be longer than a nanosecond. A long collisional time requires a low pressure. The 1420 MHz spin recoupling resonance of hydrogen atoms is too high, but the frequencies 178 MHz in the 2S$_{1/2}$ state, 59 MHz in 2P$_{1/2}$ state, and 24 MHz in 2P$_{3/2}$ are very convenient: The Coherent Raman Effect on Incoherent Light (CREIL) is this parametric transfer of energy between beams propagating in {\it excited} atomic hydrogen.

A simple computation (Appendix) shows that, in the best conditions founded on reasonable hypothesis, the frequency shift per unit of length is proportional to the inverse of the cube of the length of the light pulses. Comparing a laboratory size experiment using femtosecond lasers, with the use of ordinary light, a 10$^{15}$ longer, astronomical path is needed.

\section{Spectrum of spontaneously emitted lines of hydrogen.}\label{spec}

In a low pressure gas, excited atomic hydrogen plays two roles: 

- by its de-excitation, it emits lines, in particular the strong Lyman alpha line.

- acting as a catalyst, it allows a coherent, parametric transfer of energy between electromagnetic beams.

In the external region of the Str\"omgren sphere, the self-amplified process of creation of atoms increases quickly the density of excited atoms (graph. 1:c) until, for $r=R$, the super-radiance starts.
Inside the sphere ($r<R$), the radiance of the spontaneously emitted rays is not very large, so that these rays are not very hot, even at Lyman $\alpha$ frequency.

By the CREIL effect, the spontaneously emitted rays receive energy from the extremely hot radial rays, and give energy to the thermal background. Supposing that the source is small, seen through a small enough solid angle while the background is seen through $4\pi$ steradians, we assume (assumption A-) that the balance of energy is negative for the spontaneously emitted rays which lose energy, their frequency decreasing. Thus, the intensity of spontaneously emitted lines increases, while the path of light to exit the sphere, therefore the red-shift of previously emitted light, decreases. Thus, the intensity is maximal for the emission at $r=R$, while, at this radius the red-shift is evidently zero (graph. 2,a). The opposite assumption, (A+) may be done if the star is seen through a large solid angle.

\begin{figure}
\centerline{\includegraphics[height=8 cm]{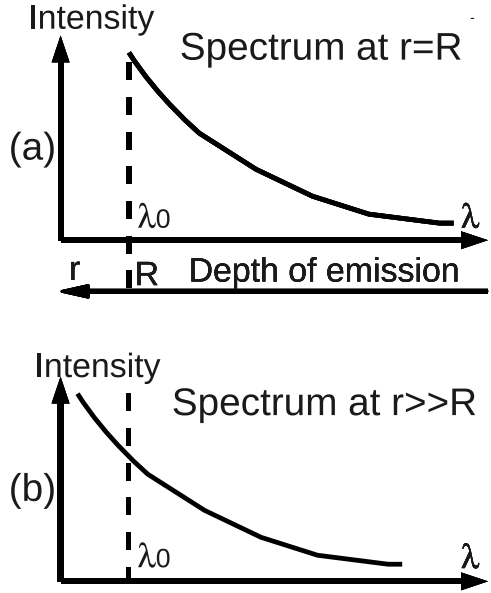}}
\caption{Theoretical spectrum of weak light emitted inside the ring, observed at distances $R$ (a) and $R1$ (b); $\lambda_0$: laboratory wavelength.}
\label{rad2}
\end{figure}

\medskip
For $r$ increasing over $R$, the radial rays transfer their energy to the tangential super-radiant beams (graph 1:e; 1:f). Thus, the mean {\it radial component} of the speed of propagation of the energy of light decreases radically, so that the density of energy of hot light (radial + amplified super-radiant) increases greatly (graph 1:h). Thus, we assume (assumption B+) that the balance of energy becomes positive for the weak, spontaneously emitted beams.

As the density of excited atoms (which produce the parametric effect) decreases fast (graph 1:c), this positive CREIL effect disappears quickly (graph 1:i). Its other parameters being nearly constant in the shell, the modulus of the CREIL shift (graph 1:i) is proportional to the product of the functions showed by graphs 1:c and 1:h.

 The broad line is shifted to higher frequencies, so that it has a high frequency intense wing, and the intensity of the red-shifted wing decreases down to zero while the red-shift increases (graph 2,b).

\medskip
We have chosen the assumptions A- and B+ which seem the more probable. The probability to have two other combinations, (A+ and B+) or (A- and B-) does not seem very large. A+ and B- is evidently impossible.

With any choice, it remains an important result: the intensity of the line is null on a side of a maximum while it decreases nearly exponentially to zero on the other side.

\section{Conclusion.}
The usual studies of the Str\"omgren systems neglect the nonlinearities of light-matter interactions, induced emissions of light and parametric interactions. Assuming that the source of light is not extremely hot and that the cloud of hydrogen is not too large, their results are however qualitatively good. A general study will lead to non-linear equations whose resolution can, now, only be done numerically with specific hypothesis. The present qualitative study shows three remarkable results in the case of an extremely hot source:

\medskip
In a large cloud of hydrogen, {\bf a small, extremely hot source is invisible}, its energy being mainly transferred to alpha lines of atomic hydrogen observed as a {\bf dotted ring} similar to TEM($l,m$) transverse modes of a laser, with a fixed value of $l$. The {\bf intensities and frequencies} of the broad lines observed inside the ring {\bf decrease simultaneously} from a blue- or red-shifted laboratory frequency.

Observed with these properties \cite{1987A,Michael}, supernova remnant 1987A seems surrounded by a Str\"omgren system distorted by inhomogeneities of the density of hydrogen. The shape of the Str\"omgren shell seems transformed to the shape of a three seeds peanut, or of an ellipsoid surrounded by a hourglass\cite{1987A}; Einstein's cross could be another application. 

\appendix

\section{Appendix A: Frequency shifts of time-incoherent light beams by coherent transfers of energy (CREIL).}
\setcounter{equation}{0}
\renewcommand{\theequation}{A{\arabic{equation}}}

To explain the wave propagation, Huygens deduced from a wave surface known at time $t$, a slightly later wave surface, at time $t+\Delta t$. For that, he supposes that each element of volume contained between wave surfaces relating to times $t-\mathrm dt$ and $t$ emits at the local speed of the waves $c$, a spherical wavelet of radius $c\Delta t$; the set of these wavelets has as envelopes the sought wave surface and a retrograde wave surface canceled by the retrograde waves emitted at other times, in particular by wave surfaces distant of $\lambda/4$.
à 1420 MHz
Recall the theory of refraction:

Let us suppose that each element of volume considered by Huygens also emits a wavelet of much lower amplitude at same frequency whose phase is delayed by $\pi/2$ (Rayleigh coherent emission). The interaction of light and matter can be modeled classically by a non resonant molecular oscillator, or by a quantum \textquotedblleft dressing\textquotedblright {} of the matter whose stationary state is mixed with other states. Except for the second order, this emission does not modify the initial amplitude. Are $E_0 \sin (\Omega T)$ the\label{shell} incident field, $E_0K\epsilon \cos(\Omega T)$ the field diffused in a layer of infinitesimal thickness $\epsilon=c\mathrm dt$ on a wave surface, and $K$ a coefficient of diffusion; the total field is:

\begin{equation}
E=E_0[\sin(\Omega t)+K\epsilon \cos(\Omega t)]\label{refr}
\end{equation}
\begin{equation}
\approx E_0[\sin(\Omega t)\cos(K\epsilon)+\sin(K\epsilon )\cos(\Omega t)]=E_0\sin(\Omega t -K\epsilon).
\end{equation}
 This result defines the index of refraction $n$ by the identification 
\begin{equation}
K=2\pi n/\lambda=\Omega n/c.\label{indice}
\end{equation}

\medskip
The previous theory shows that refraction results from the coherent Rayleigh scattering. Replace the coherent Rayleigh scattering  by a coherent Raman scattering. The frequency of the Raman scattered light is shifted by $\omega$, but this light has no initial phase shift.

Setting $\omega$ the Raman frequency shift, $K'>0$ the anti-Stokes diffusion coefficient, formula \ref{refr} becomes:
\begin{equation}
E=E_0[(1-K'\epsilon)\sin(\Omega t)+K'\epsilon \sin((\Omega+\omega)t)].
\end{equation}
In this equation, incident amplitude is reduced to obtain the balance of energy for $\omega=0$.
\begin{equation}
E=E_0\{(1-K'\epsilon)\sin(\Omega t)+K'\epsilon[\sin(\Omega t)\cos(\omega t)+\sin(\omega t)\cos(\Omega t)]\}.
\end{equation}
$K'\epsilon$ is infinitesimal; suppose that between the beginning of a pulse at $t=0$ and its end, $\omega t$ is small; the second term cancels with the third, and the last one transforms:
\begin{eqnarray}
E\approx E_0[\sin\Omega t+\sin(K'\epsilon\omega t)\cos(\Omega t)]\nonumber\\
E\approx E_0[\sin(\Omega t)\cos(K'\epsilon\omega t)+ \sin(K'\epsilon\omega t)\cos(\Omega t)=E_0\sin[(\Omega+K'\epsilon\omega)t].\label{eq4}
\end{eqnarray}
Hypothesis $\omega t$ small requires that Raman period $2\pi/\omega$ is large in comparison with the duration of the light pulses; to avoid large perturbations by collisions, the collision-time must be larger than this duration.  This is a particular case of the condition of space coherence and constructive interference written by Lamb.

Stokes contribution, obtained replacing $K'$ by a negative $K''$, must be added. Assuming that the gas is at equilibrium at temperature $T$, $K'+K''$ is proportional to the difference of populations in Raman levels, that is to $\exp[-h\omega/(2\pi kT)]-1 \propto \omega/T$.

$K'$ and $K''$ obey a relation similar to relation \ref{indice}, where Raman polarisability which replaces the index of refraction is also proportional to the pressure of the gas $P$ and does not depend much on the frequency if the atoms are far from resonances; thus, $K'$ and $K''$ are proportional to $P\Omega$, and $(K'+K'')$ to $P\Omega \omega/T$. Therefore, for a given medium, the frequency shift is:

\begin{equation}
\Delta\Omega=(K'+K'')\epsilon\omega\propto P\epsilon\Omega\omega^2/T.\label{delom}
\label{red-shift}
\end{equation}

The relative frequency shift $\Delta\Omega/\Omega$ of this space-Coherent Raman Effect on time-Incoherent Light (CREIL) is nearly independent on $\Omega$ and proportional to the integral of $Pc{\rm d}t$, that is to the column density of active gas along the path \cite{Mor98b,Moret}.

The path needed for a given (observable) red-shift is inversely proportional to $P\omega^2$. At a given temperature, assuming that the polarisability does not depend on the frequency, and that $P$ and $\omega$ may be chosen as large as allowed by Lamb's condition, this path is inversely proportional to the cube of the length of the pulses: an observation easy in a laboratory with femtosecond pulses requires astronomical paths with ordinary incoherent light.

\end{document}